\documentclass[preprint,
prl,
]{revtex4-2}

\usepackage{layout}
\usepackage{graphicx,color}
\usepackage{amsmath,amsfonts,amssymb}
\usepackage{bm}
\usepackage[symbol]{footmisc}
\usepackage{epstopdf}
\usepackage[section]{placeins}
\usepackage[font={footnotesize}]{caption}
\usepackage{ulem}

\newcommand{\ju}[1]{{\color{magenta}#1}}

\newcommand{\nn}{\nonumber}

\begin{document}

\title[Self-sustained oscillations of active viscoelastic matter]{Self-sustained oscillations of active viscoelastic matter}

\author{Emmanuel L. C. VI M. Plan}
\affiliation{Institute of Theoretical and Applied Research, Duy Tan University, Ha Noi 100 000, Viet Nam}
\affiliation{Faculty of Natural Science, Duy Tan University, Da Nang 550 000, Viet Nam}

\author{Huong Le Thi}
\affiliation{Department of Mathematics and Informatics, Thang Long University, Ha Noi 100 000, Viet Nam}

\author{Julia M. Yeomans}
\affiliation{The Rudolf Peierls Centre for Theoretical Physics, Department of Physics, University of Oxford,\\ Clarendon Laboratory, Oxford OX1 3PU, United Kingdom}

\author{Amin Doostmohammadi}
\affiliation{The Niels Bohr Institute, University of Copenhagen, Blegdamsvej 17, 2100 Copenhagen, Denmark}

\begin{abstract}
Models of active nematics in biological systems normally require complexity arising from the hydrodynamics involved at the microscopic level as well as the viscoelastic nature of the system. Here we show that a minimal, space-independent, model based on the temporal alignment of active and polymeric particles provides an avenue to predict and study their coupled dynamics within the framework of dynamical systems. In particular, we examine, using analytical and numerical methods, how such a simple model can display self-sustained oscillations in an activity-driven viscoelastic shear flow.
\end{abstract}

\maketitle


\section{Introduction}
Active nematics are a class of living or synthetic materials that possess some form of orientational order and that  are continuously driven by their constituent living or artificially active elements~\cite{Marchetti13,doostmohammadi2018active}. The orientational order could arise from the elongated shape of the particles, such as in filamentous cytoskeletal elements~\cite{sanchez2012,kumar2018tunable,maroudas2020topological}, rod-like bacteria~\cite{volfson2008biomechanical,copenhagen2020bacteria,meacock2020bacteria} or spindle-shaped fibroblasts~\cite{duclos2018spontaneous} and neural progenitor stem cells~\cite{kawaguchi2017topological}, or could be an emergent feature of deformable particles such as in monolayers of epithelial cells~\cite{saw2017,blanch2018turbulent}. Whether it is moving bacteria or eukaryotic cells, or subcellular filaments that are put in motion by motor proteins, the common feature of all these systems is the input of energy at the particle level, that governs system-wide patterns of motion and self-organisation in the active material. While active nematic models have been indispensable in characterising various systems of active matter, their inherent complexity, that requires coupling between density, velocity, and the orientational field, still limits theoretical advances and often requires extensive numerical computations in order to obtain generic and fundamental rules of dynamical self-organisation in active matter.

To add to this complexity, most realisations of active matter do not simply exist in homogeneous and isotropic environments. Rather, the microenvironment surrounding living materials is often characterised by its own complex structure and responses that can feedback to the active matter dynamics. Prominent among these is viscoelasticity that allows living materials to constantly deform their surroundings and then to exploit the deformation and relaxation dynamics of their viscoelastic environment as an external spatio-temporal input for their own self-organisation. Examples include bacterial biofilm formation in extracellular matrices~\cite{Nadel2015,Laura2015,vidakovic2018dynamic}, epithelial and fibroblast cells self-propelling within interconnected networks of collagen~\cite{Shawn2017,Jared2018,chaudhuri2020effects}, and subcellular filaments self-organising within the crowded environment of the cytoplasm~\cite{fletcher2010cell,pritchard2014mechanics}. Recent theoretical and experimental studies have revealed the emergence of various forms of collective behavior for active particles suspended within - or in contact with - a viscoelastic environment. In particular, numerical models of active viscoelastic nematics~\cite{Hemingway2016} and active nematics in contact with viscoelastic surroundings~\cite{emmanuel2020active} have demonstrated the impact of polymeric stresses and viscoelastic relaxation times on active nematic patterns in confined geometries and in models of cell self-propulsion and cell division. 
Oscillations have been shown using a full space-dependent hydrodynamic models of active matter which includes viscoelasticity ~\cite{Hemingway2016}.
Moreover, combined experiments and theory have recently shown the emergence of directional switching in bacterial collectives suspended within a viscoelastic medium and confined in circular geometries~\cite{songliu2021visco}. 
Consistent with the experiments, a simplified model that maps into a Fitzhugh-Nagumo oscillator was  deduced~\cite{songliu2021visco}.

Here we show that an alternative model of active viscoelastic matter relying only on the orientation of both active and viscoelastic components along with their respective relaxation timescales can also produce oscillatory dynamics. 
This model was first introduced in a recent paper~\cite{emmanuel2021} for active matter in contact with viscoelastic surroundings to reproduce numerical observations of the reversals of the flow field in a nematic driven by pulsatile activity.
We explore the generic dynamics of this minimal, space-independent model within the framework of dynamical systems, and prove analytically that sustained oscillatory behaviour can arise spontaneously as an effect of the viscoelastic feedback to the active forcing.
We provide a phase diagram for the emergence of the oscillations in terms of the dominant time scales of the system, and differentiate their form from sinusoidal oscillations.
The equations we consider are not algebraic and cannot be naively truncated nor mapped to well-known oscillators.
Also, we remark that the physics described in this work is entirely different from that seen in the oscillations in purely active models~\cite{Giomi2011,Woodhouse2012}, because of the presence here of a polymeric component.

\section{Model}
Consider an active nematogen and a polymer, oriented at angles $\theta_n$ and  $\theta_p$ respectively with respect to the $x$-axis. Then, modelling both the nematogen and the polymer as rigid rods rotating in 2D in the shear $\dot{\gamma}$ created by the active stresses and the polymer relaxation, we consider the simplified, space-independent, equations of motion (first proposed in~\cite{emmanuel2021})
\begin{eqnarray}
\dot{\theta}_n&=&\dot{\gamma}\lambda_n\cos2\theta_n-(\theta_n/t_n), \label{eq:eqn}\\
\dot{\theta}_p&=&\dot{\gamma}\lambda_p\cos2\theta_p-(\theta_p/t_p), \label{eq:eqp}\\
\dot{\gamma}&=&\frac{\zeta}{4\nu}\sin2\theta_n-\frac{\nu_p}{4\nu t_p}\sin2\theta_p. \label{eq:shearrate}
\end{eqnarray} 
The first terms on the right-hand side of Eqs.~(\ref{eq:eqn}) and~(\ref{eq:eqp}) describe the nematic and polymeric response to the shear,  $\dot{\gamma}$, in the limit that vorticity can be neglected. $\lambda_n,\lambda_p>0$ are phenomenological parameters related to the aligning/tumbling properties, see for example~\cite{aigouy2010cell}.  The second terms model exponential relaxation to equilibrium, with $t_n, t_p$  the relaxation timescales of the nematic and polymer. The shear rate, Eqn.~(\ref{eq:shearrate}), is the off-diagonal term in the stress balance $\Pi_{\rm visc}=\Pi_{\rm active}+\Pi_{\rm polymer}$, where $\nu$ is the fluid viscosity and $\nu_p$ is the polymeric contribution to viscosity~\cite{emmanuel2021}. The dynamics of the nematogen and polymer are coupled through the shear which describes the balance between the extensile ($\zeta>0$) stress resulting from the activity of the active  nematic and the contractile stress created by the polymer relaxation.

These equations represent a simplification motivated by the full hydrodynamic model investigated numerically in~\cite{emmanuel2021} which showed that activity pulses in a system interacting with a viscoelastic environment can spontaneously generate flow reversals driven by the stress built up in the surrounding, passive, medium. In particular, replacing the space-dependent flow by a uniform shear results in only two dynamical variables, which allows us to fully investigate the phase space and to prove the existence of oscillations.

\section{Results and Discussion}
We discuss the results in two steps: first we will fix the value of the polymer viscosity $\nu_p$ and then we keep the elastic modulus, i.e., the ratio of the polymer viscosity to the polymer relaxation time $A_p=\nu_p/t_p$ constant.

\subsection{Fixed polymer viscosity $\nu_p$}
We first consider the case when polymer contribution to viscosity $\nu_p$ is fixed and  the elastic modulus $A_p$ varies inversely with $t_p$.
Note that in the absence of activity ($\zeta=0$) the system relaxes to the equilibrium and trivial fixed point $\theta_n=\theta_p=0$. 
Moreover, for finite activity, in the limit of infinite relaxation times, only the active force dictates the shear and thus the steady-state solution is $\theta_n=\pm\pi/4, \theta_p=\pm\pi/4$.   
Henceforth, without losing generality we restrict our analysis of the system to the box $\Omega=[-\pi/4,\pi/4]^2$. 
This two-dimensional system $\bm\theta^T=(\theta_n,\theta_p)$ can then be written as $\dot{\bm\theta}=\bm f(\bm\theta)$, with $\bm f:\mathbb{R}^2\rightarrow \mathbb{R}^2$. By introducing the time scale of active stress generation $t_a=\nu/|\zeta|$, Eqs.~\eqref{eq:eqn},\eqref{eq:eqp} can be expressed in terms of dimensionless numbers $T_p=t_p/t_a$ and $T_n=t_n/t_a$:
\begin{subequations}
\begin{align}
\dot{\theta}_n&=f_n(\theta_n,\theta_p) \nn\\
&=\frac{1}{t_a}\left(\frac{\lambda_n}{4}\cos2\theta_n\sin2\theta_n-\frac{\lambda_n\nu_p}{4\nu T_p}\cos2\theta_n\sin2\theta_p-\frac{\theta_n}{T_n}\right), \hfill\\
\dot{\theta}_p&=f_p(\theta_n,\theta_p)\nn\\
&=\frac{1}{t_a}\left(\frac{\lambda_p}{4}\cos2\theta_p\sin2\theta_n-\frac{\lambda_p\nu_p}{4\nu T_p}\cos2\theta_p\sin2\theta_p-\frac{\theta_p}{T_p}\right).
\end{align}
\label{eq:dimensionless}
\end{subequations}
From Eqs.~\eqref{eq:dimensionless}, it is easy to see that the active timescale modulates the speed of the evolution of the system towards any asymptotic limit, with larger $t_a$ meaning slower dynamics.

To examine long-term dynamics, a linear stability analysis is performed by examining at any fixed point $(\theta_n^*,\theta_p^*)$ the eigenvalues $e_i=(\operatorname{Tr}J\pm\sqrt{(\operatorname{Tr}J)^2-4\operatorname{Det}J})/2,$ where $J_{ij}=\partial f_i/\partial \theta_j$ is the Jacobian. 
The origin $(0,0)$ is a trivial fixed point for all values of the parameters, and a short calculation shows that a necessary condition for stability is $\operatorname{Tr}J=\frac{\lambda_n}{2}-\frac{1}{T_n}-\frac{\lambda_p\nu_p}{2\nu T_p}-\frac{1}{T_p}<0$. For simplicity, we assume $\lambda_n=\lambda_p=\nu=\nu_p=1$ for the rest of this work. The condition for stability then simplifies to $T_p<3T_n/(T_n-2)$.

\begin{figure}
    \centering
    \includegraphics[width=0.49\textwidth]{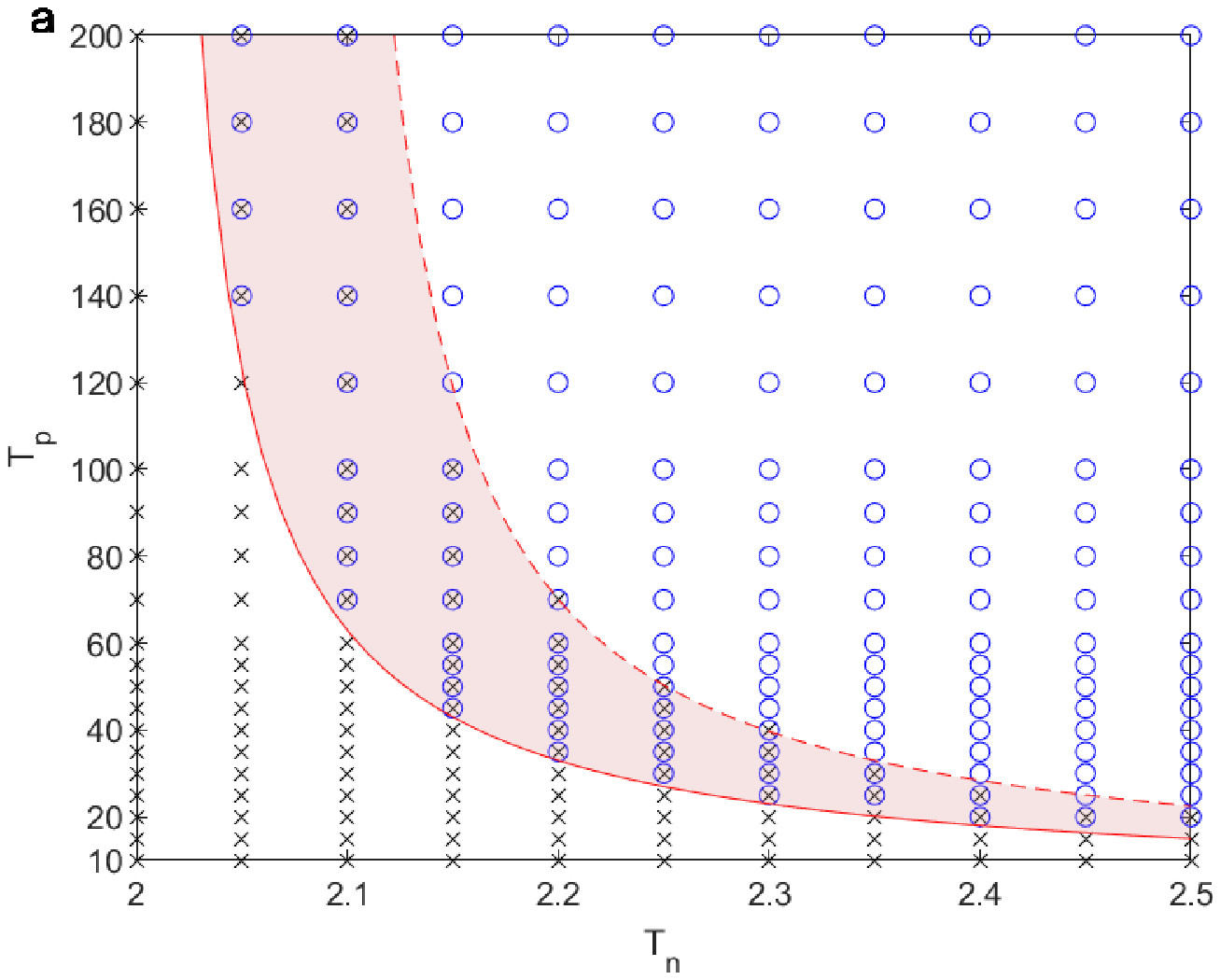}
    \includegraphics[width=0.49\textwidth]{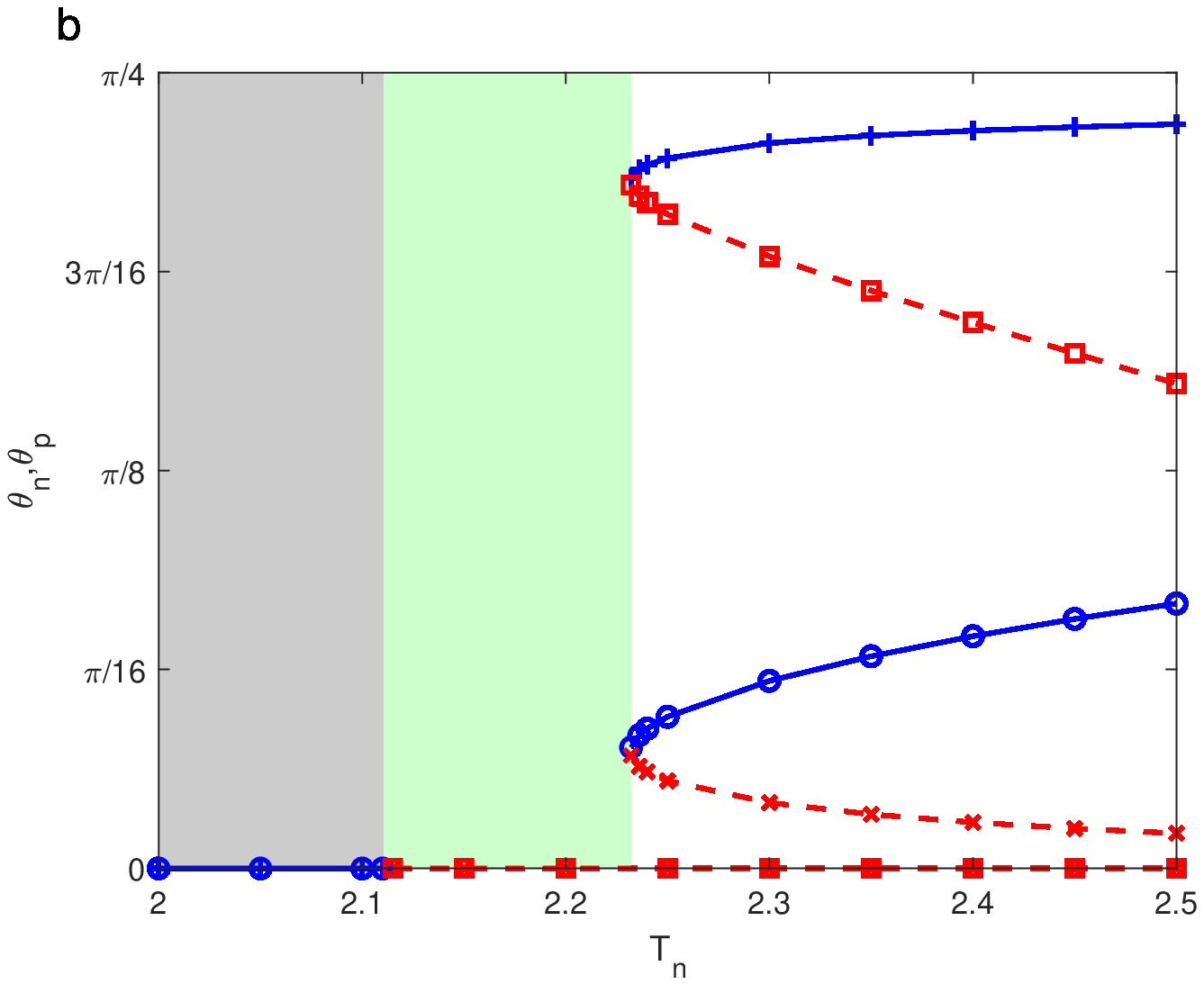}
    \caption{(a) Phase diagram in the $T_p$-$T_n$ space for $\lambda_n=\lambda_p=\nu=1$. Crosses ($\times$) denote the region where the only fixed point is the trivial one, whether stable or unstable. Circles ($\circ$) denote the region where the trivial fixed point is unstable. The red shaded region is where oscillations occur. The solid red line is $T_p=3T_n/(T_n-2)$ and the dashed line is a numerical estimate of the second border of the oscillatory region. 
    (b) Bifurcation diagram showing $\theta_n (\theta_p)$ as a function of $T_n$, with $T_p=60$. Stable points are denoted by circles ($+$ marks) in blue and unstable points are denoted by cross (square) marks in red. The gray-shaded region ($T_n\lesssim2.11$) is where the trivial fixed point is stable, green is where oscillatory behavior occurs, and the white region ($T_n\gtrsim2.23$) is where more than one fixed point is present. The fixed points with negative values of $\theta_n,\theta_p$ will be a reflection of the current figure with respect to the axis.}
    \label{fig:phasediag}
\end{figure}

Thus any trajectory that begins in the neighborhood of the origin either asymptotically approaches this point if it is stable, or wanders away from it if $\operatorname{Tr}J>0$. This is indeed the case, as is illustrated in a typical phase diagram in Fig.~\ref{fig:phasediag}a, where the red solid line
gives the line $T_p=3T_n/(T_n-2)$ and the blue circles above it denote the region where a trajectory is \textit{not} attracted to the trivial fixed point. If the trivial fixed point is attractive (points below the red line), the angles $\theta_n,\theta_p$ asymptotically align towards 0, possibly in an oscillatory motion.
Physically speaking, small values of either $T_n$ or $T_p$ imply rapid orientational relaxation dominating over the  active forces.

Aside from the trivial fixed point, the mix of algebraic and transcendental functions in $\bm f$ prevents us from analytically obtaining other fixed points, unlike other relaxation oscillators in literature (e.g. the Van der Pol equation)\cite{Strogatz1994}. We thus approach the problem numerically and obtain both the location and the stability of any other fixed points of this system, as demonstrated in the bifurcation diagram in Fig.~\ref{fig:phasediag}b. 
To obtain this diagram trajectories were initiated within the entire space and allowed to evolve until convergence to unique fixed points. The locations of these fixed points can be anticipated by drawing the nullclines $\dot{\theta}_n=\dot{\theta}_p=0$ for different values of the parameters and dimensionless values.
A similar diagram can be constructed for when the bifurcation parameter is $T_p$.

There are three distinct regions in this figure. The first region, when the dimensionless nematic relaxation time, $T_n$, is sufficiently small (gray region), corresponds to when the trivial fixed point is stable. Here there are no other fixed points, and the origin provides the global attracting point of the system. The trivial fixed point loses its stability with increasing $T_n$.
By contrast, if $T_n$ is sufficiently large (white region), a saddle node bifurcation occurs at the intersection of the nullclines $\dot{\theta}_n=\dot{\theta}_p=0$ and spontaneously gives birth to a pair of stable and unstable nodes~\cite{Strogatz1994}. The stable node approaches $(\pm \pi/4,\pm \pi/4)$ as $T_n\rightarrow\infty,$ as expected from our earlier arguments. \\

For a narrow intermediate range of the dimensionless number $T_n$ (the green region) the only fixed point is that at the origin, which is now unstable. We show that in this region the dynamical system undergoes an oscillatory motion in the nematic and polymer alignment angles $\theta_n,\theta_p$. To do this, we will utilize the Poincar\'e-Bendixson Theorem~\cite{Strogatz1994,Lawrence2001}, which can be invoked given a two-dimensional, continuously differentiable differential equation defined on an open set encompassing a compact subset $\Omega^*$ that contains a trajectory confined within $\Omega^*$. The Poincar\'e-Bendixson Theorem asserts that if $\Omega^*$ does not contain any fixed points, then $\Omega^*$ contains a closed orbit. In our case, this closed orbit corresponds to an oscillatory motion in the time evolution of the angles.

\begin{figure}
    \centering
    \includegraphics[width=0.49\textwidth]{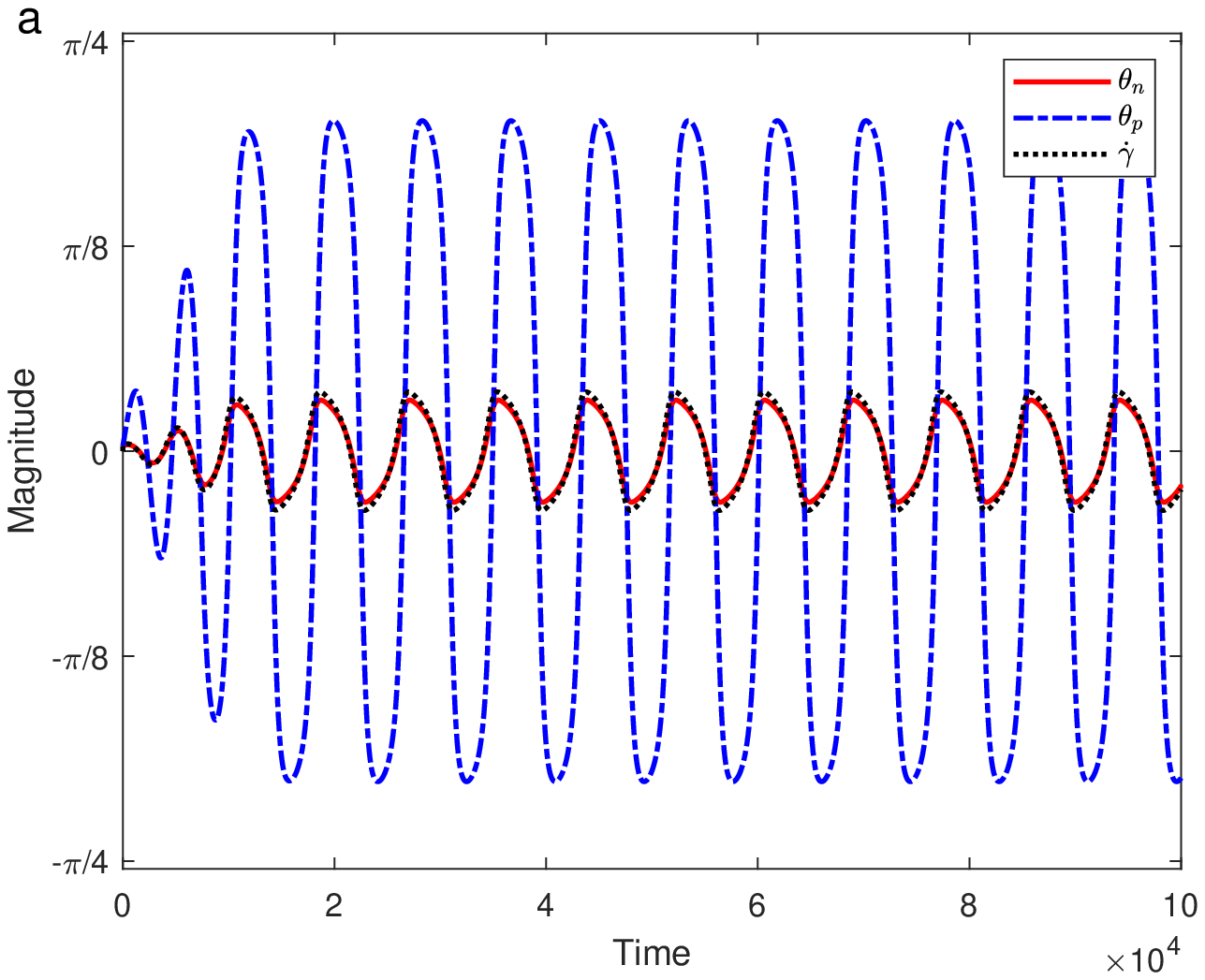}
    \includegraphics[width=0.49\textwidth]{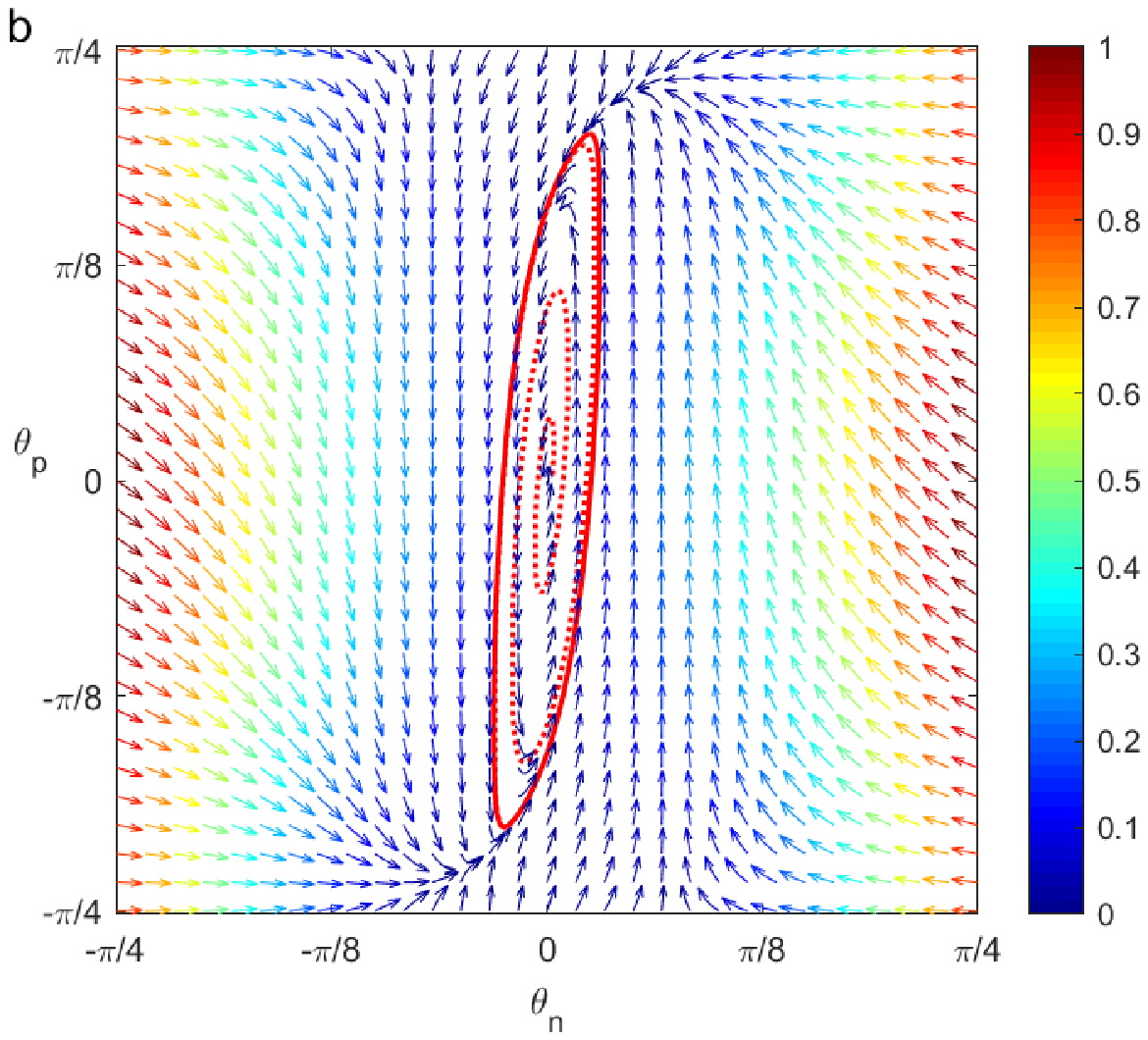}
    \caption{(a) Time series displaying oscillatory motion for $T_n=2.2, T_p=60, t_a=40$. (b) Corresponding vector plot and trajectory (red dotted line) of Eqs.~\eqref{eq:dimensionless} in the $\theta_n$-$\theta_p$ space. Note that the vectors in the figure are rescaled to the maximum magnitude in the plot using the Matlab visualization package Streak \cite{Bertrand2021streak}.}
    \label{fig:oscillations}
\end{figure}

Note that the $f_i$ as defined in Eqs.~\eqref{eq:dimensionless} are continuous in $\mathbb{R}^2\supset\Omega$ and it can be easily shown (see Appendix) that $\Omega$ is a trapping region, i.e. all trajectories at the boundary point towards the interior of $\Omega$, confining all trajectories within $\Omega$. By erecting a small neighborhood $\mathcal{B}$ of radius $\epsilon$ around the repeller $(0,0)$, then $\Omega^*=\Omega\setminus\mathcal{B}(\bm 0;\epsilon)$ is a compact trapping region. As a consequence of the Poincar\'e-Bendixson Theorem, there is a closed orbit in this region, which implies oscillatory motion in $\theta_n,\theta_p$, as exemplified in the time series in Fig.~\ref{fig:oscillations}a and the red orbit in Fig.~\ref{fig:oscillations}b. The vector field in this figure further supports the attractive nature of this limit cycle. 

Note that the bifurcation diagram in Fig.~\ref{fig:phasediag}b is simply a slice of the red region in the full phase diagram in Fig.~\ref{fig:phasediag}a, taken at a dimensionless polymer relaxation time, $T_p=60$. Indeed it is in this red region of the phase diagram, corresponding to only one fixed point which is unstable, where the conditions of the Poincar\'e-Bendixson Theorem are satisfied. The transition from an attractive fixed point into a limit cycle shows that the curve $T_p=3T_n/(T_n-2)$ marks a supercritical Hopf bifucation.

\subsection{Fixed elastic modulus $A_p$}
Next we study the case with fixed elastic modulus $A_p=\nu_p/t_p$. In contrast to the Eqs.~\eqref{eq:dimensionless} above, the dimensionless relaxation variable $T_p$ ceases to appear in the shear terms, and there is a less simple dependence on the active timescale $t_a$. In this case, Eqs.~\eqref{eq:eqn}-\eqref{eq:shearrate} become
\begin{subequations}
\begin{align}
\dot{\theta}_n&=\frac{1}{t_a}\left(\frac{\lambda_n}{4}\cos2\theta_n\sin2\theta_n-\frac{\lambda_n A_p t_a}{4\nu }\cos2\theta_n\sin2\theta_p-\frac{\theta_n}{T_n}\right), \hfill\\
\dot{\theta}_p&=\frac{1}{t_a}\left(\frac{\lambda_p}{4}\cos2\theta_p\sin2\theta_n-\frac{\lambda_p A_p t_a}{4\nu}\cos2\theta_p\sin2\theta_p-\frac{\theta_p}{T_p}\right).
\end{align}
\label{eq:fixedmod}
\end{subequations}
As in the previous case of constant polymer viscosity, the origin is a fixed point for all values of  $T_n,T_p$. However, in the limit of vanishing relaxation $T_n,T_p\rightarrow\infty$, all the points in the curve $4\dot{\gamma}=\sin2\theta_n-A_p\sin2\theta_p/\nu=0$ are fixed points of the system, and the  domain to be considered cannot a priori be limited to $\Omega$. Nevertheless, in the finite $T_n,T_p$ case, it will be possible to restrict the analysis to $\Omega$ and to prove that oscillatory motion may still occur.

\begin{figure}
    \centering
    \includegraphics[width=0.49\textwidth]{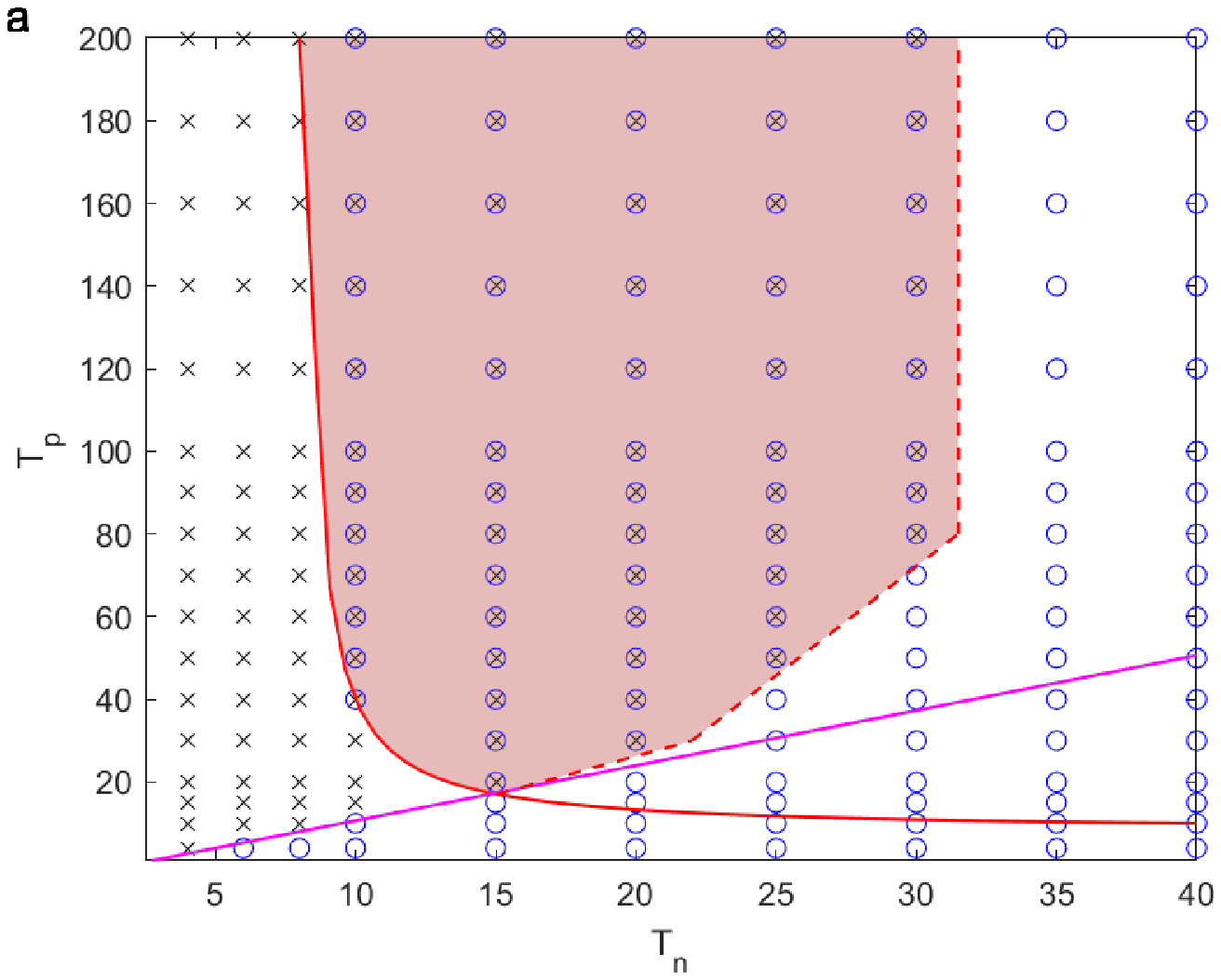}
    \includegraphics[width=0.49\textwidth]{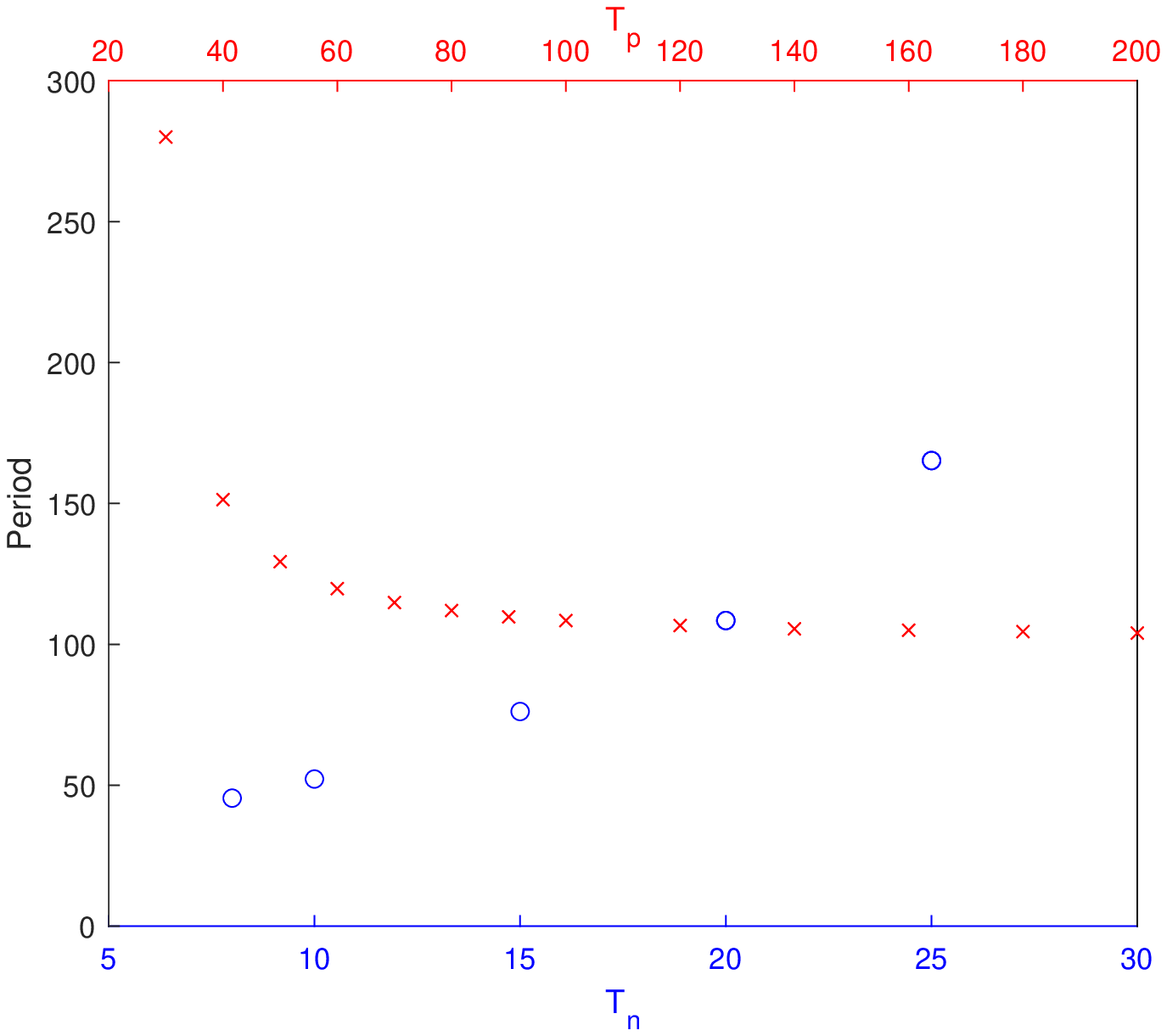}
    \caption{(a) Phase diagram for $A_p=0.5, t_a=1.5$ for $\lambda_n=\lambda_p=\nu=1$. The red curve is $\operatorname{Tr}J=0$ and the pink curve is $\operatorname{Det}J=0$. Crosses ($\times$) denote the region where the only fixed point is the trivial one, whether stable or unstable. Circles ($\circ$) denote the region where the trivial fixed point is unstable. The red shaded region is where oscillations occur.
The dashed line is an approximate estimate of the boundary demarcating the region displaying oscillations. (b)  The oscillation period for the active time scale $t_a=1.5$ as a function of $T_n$ (blue, with $T_p=100$) or $T_p$ (red, with $T_n=20$).}
    \label{fig:fixedelastic}
\end{figure}

Performing the same stability analysis at the origin (with finite $T_n,T_p$), we now require two necessary conditions to establish stability: $\operatorname{Tr} J<0$ and $\operatorname{Det} J>0$, which are both dependent on the values of $A_p$ and $t_a$. As before $0>\operatorname{Tr} J=\frac{\lambda_n}{2}-\frac{1}{T_n}-\frac{\lambda_p A_p t_a}{2\nu}-\frac{1}{T_p}$ 
which simplifies to $\frac{1}{T_n}+\frac{1}{T_p}>\frac{1-A_p t_a}{2}$ if $\lambda_n=\lambda_p=\nu=1$. Note that if $A_p t_a\geq1$, the origin remains stable for all values of $T_n,T_p$, which implies that either the elastic modulus or the active time scale (or both) must be small to ensure stability. 
The other necessary condition for stability ($\operatorname{Det} J>0$), for the case  $\lambda_n=\lambda_p=\nu=1$, and
under the assumption that $T_n>2$, is $T_p>\frac{T_n-2}{A_p t_a}.$ This condition is shown as a pink line in Fig.~\ref{fig:fixedelastic} in a typical $T_n$-$T_p$ phase space when $A_p t_a<1$. Together the conditions now clearly dictate the boundary of the stable region.

We numerically explore different combinations of $A_p t_a$ and confirm that attractive orbits still exist as was found in the fixed $\nu_p$ case. As before, oscillations occur when the trivial fixed point is unstable and when no other singularity is present in the domain $\Omega$, as seen in the typical phase diagram for $A_p t_a<1$ in Fig.~\ref{fig:fixedelastic}, although oscillations are observed for a possibly wider range $k_1<T_n<k_2$, where $k_1,k_2$ are dependent on $A_p t_a$. The nematic relaxation $t_n$ must remain sufficiently longer than $t_a$, else the activity-induced shear dominates suppressing the oscillations. 
The values of $T_p$ that permit oscillations, in contrast to the fixed polymer viscosity case, can be much larger, owing to the fact that polymer stress in Eq.~\eqref{eq:shearrate} is now independent of $T_p$.

Note that it would not be possible to guarantee a trapping region in the entire domain of $\theta_n,\theta_p$, and the presence of several fixed points will anyway prevent the use of the Poincar\'{e}-Bendixson Theorem. However the Poincar\'{e}-Bendixson Theorem can again be used if we restrict the initial conditions \ju{to} a smaller domain, e.g. to $\Omega$. The resulting vector diagrams and bifurcation diagrams are similar to the previous case. We also note that oscillations occur right before the bifurcations appear  \textit{spontaneously} in the bifurcation diagram, rather than the origin \textit{splitting} into a pair of fixed points. The calculations to establish the trapping region, needed to utilise the Poincar\'{e}-Bendixson Theorem, are exactly the same as in the previous case, thanks to the same terms vanishing.
\\

\subsection{Periodicity}
Previous models and experiments have shown that the oscillations of active matter in viscoelastic environments exhibit non-sinusoidal trajectories, consistent with the behaviour of dynamical systems as they move farther from a bifurcation point~\cite{songliu2021visco}. Here we characterize the oscillatory motion of our model and  confirm that a similar behaviour can be observed for the periodicity and the shape of the orbits for different values of $T_n$ and $T_p$. 

\begin{figure}
    \centering
    \includegraphics[width=0.49\textwidth]{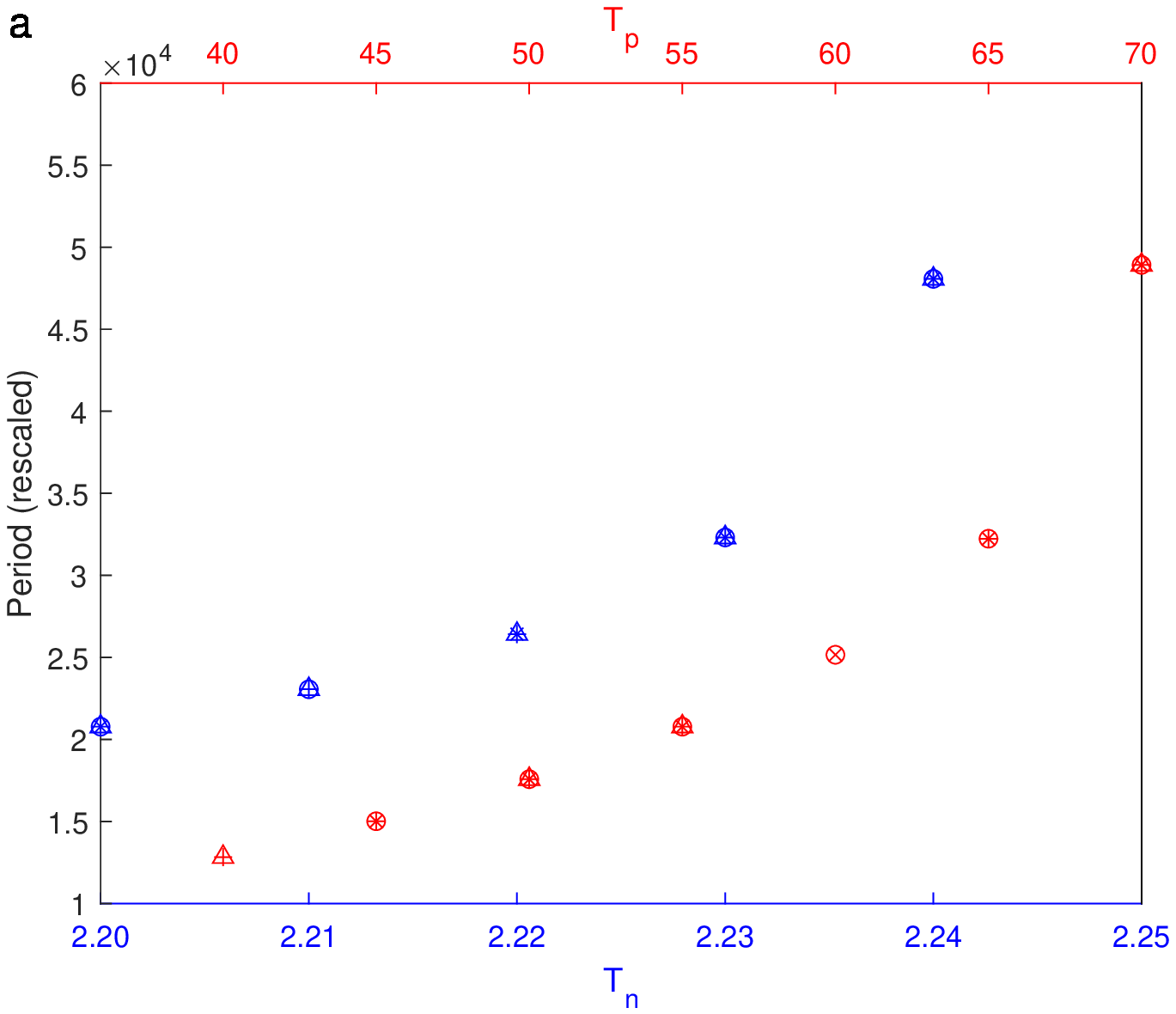}
    \includegraphics[width=0.49\textwidth]{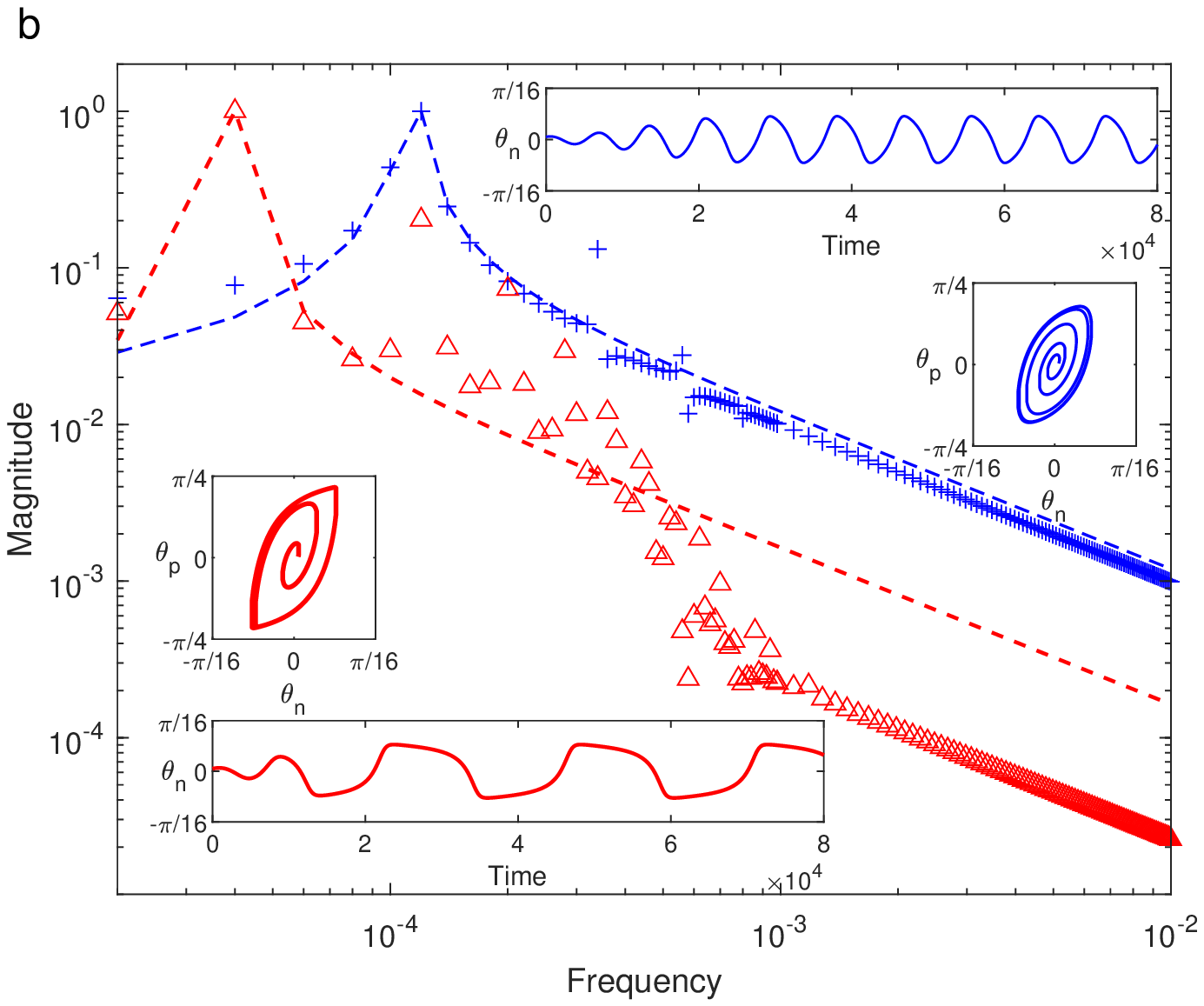}
    \caption{ (a) The oscillation period for the fixed polymer viscosity case, with active time scale $t_a=120$ and as a function of $T_n$ (blue, with $T_p=55$) or $T_p$ (red, with $T_n=2.2$), obtained by rescaling simulations from $t_a=20 (\triangle)$, $t_a=30 (+)$, $t_a=40(\times)$, $t_a=60 (\circ)$. (b) Points indicate the normalized discrete Fourier transform of $\theta_n$ as a function of frequency; the dashed lines show the sinusoidal signal of corresponding periodicity (with an exponential tail arising from the numerical approximation). The parameters are $t_a=60,T_n=2.2$ and $T_p=50$ (blue $+$), and $T_p=70$ (red $\triangle$), and the values are normalized for each case by the maximum value of the signal. The insets indicate the corresponding time series of $\theta_n$ and orbits in the $\theta_n$-$\theta_p$ space.}
    \label{fig:period}
\end{figure}

We first report results for fixed polymer viscosity.
The period of the oscillatory behavior, calculated numerically, increases nonlinearly as either $T_n$ and $T_p$ increases (Fig.~\ref{fig:period}a). The active timescale $t_a$ simply rescales the period for a fixed $T_n$-$T_p$ pair, as evidenced by the perfect collapse of the measured periods.
As expected, the increase in the period is accompanied by a decreasingly sinusoidal shape in the time series of the angles. For instance, the insets at the top and bottom of Fig.~\ref{fig:period}b illustrate the evolution of the nematic alignment angle $\theta_n$ for identical values of $T_n$. The blue curve ($T_p=50$) is barely indistinguishable from a sinusoidal curve whereas the red curve ($T_p=70$) is reminiscent of the Fitzhugh-Nagumo or the Van der Pol oscillators, which are characterised by alternating slow-fast relaxation~\cite{Fitzhugh1955,Strogatz1994}.

To quantify the deviation of these trajectories from a sinusoid, we numerically calculate their Fourier spectra and compare them to the spectra of sinusoids with corresponding periods (Fig.~\ref{fig:period}b), with the peak in the frequencies indicating the inverse of the period. Oscillations that result from Eqs.~\eqref{eq:dimensionless} (in markers) indicate that the frequency distributions differ from the discretized sinusoidal counterparts (dashed lines) with more significantly-activated frequencies as $T_p$ or $T_n$ increases. This irregularity can be visually corroborated by looking at the orbits in the $\theta_n$-$\theta_p$ phase space as provided in the accompanying insets. A sinusoidal trajectory would correspond to an elliptical shape, whereas the slow-fast dynamics manifests as a rugby-ball-shaped orbit, with nearly pointed tips. Moreover, we remark that the behaviour is closer to sinusoidal for pairs $T_n$-$T_p$ near to the supercritical Hopf bifurcation (where $\operatorname{Tr} J=0$, red solid line in Fig.~\ref{fig:phasediag}a), consistent with the findings in \cite{songliu2021visco}.

For the case of fixed elastic modulus, similar non-sinusoidal orbits are observed. While the period increases with $T_n$ like in the fixed $\nu_p$ case, it decreases with $T_p$ and approaches a fixed value in this case (Fig.~\ref{fig:fixedelastic}b). This asymptotic decrease in period is attributed to the polymer orientation being dominated by the fluctuating shear, which is independent of $T_p$ in the fixed elastic modulus case. Slower polymer relaxation results in the slow-fast transition being attained faster.\\

\section{Conclusion}
Our results help to elucidate the physics that underlies the existence of oscillatory dynamics.  We first note that our minimal model of an active viscoelastic fluid displays oscillatory behavior when $T_n<T_p$. As such, the nematic director relaxes and seemingly equilibrates faster than the polymer, but polymer orientation attains larger values because of the stronger tendency to align to the shear (weaker relaxation term). Since the shear rate is an intricate balance between the active and polymeric stresses, then as the nematic angle  approaches its (non-interacting) steady state, it gets pushed away from this \textit{equilibrium} by feedback arising from the slower but stronger polymer force. Indeed, if $\theta_p$ is sufficiently larger than $\theta_n$, then $\sin 2\theta_p>(T_p\nu/\nu_p) \sin 2\theta_n$ (for fixed polymer viscosity, $\nu_p$) or $\sin 2\theta_p>(\nu/A_p t_a) \sin 2\theta_n$ (for fixed elastic modulus, $A_p$) thereby creating a polymeric stress large enough to flip the sign of the shear, $\dot{\gamma}.$ The nematic angle $\theta_n$ then starts to realign to the new direction and overshoots beyond 0, flipping the sign of the active stress and the intensifying the shear rate in that opposite direction, resulting in a slow-fast dynamics.

To summarise, in this paper, we show how the viscoelastic components of a fluid can feedback on the shear flow generated by an active component and result in self-sustained oscillatory dynamics. The oscillations occur for specific ranges of dimensionless numbers relating the relaxation timescales of the active and viscoelastic components with the active timescale. 
In this range of values, linear stability analysis shows that the trivial fixed point of the system loses its stability, while numerical simulations verify that no other fixed point exists in the domain spanned by the orientation angles of the nematic and polymer, $\theta_n$ and $\theta_p$, respectively. 
Whether the polymer viscosity or the elastic modulus is fixed, the Poincar\'{e}-Bendixson Theorem can then be used to establish the existence of a limit cycle in (a subset of) the two-dimensional phase plane and, consequently, the oscillatory behavior in the shear rate and flow direction. 
For the case where the elastic modulus $A_p$ is kept constant, further constraints on the value of the product $A_p t_a$ are necessary to observe the oscillatory motion.
The period for a full cycle increases and the oscillations become less sinusoidal as either the nematic or the polymer relaxation becomes slower.
A slow-fast dynamics is observed, with the faster dynamics occurring once $\theta_n$ has flipped its orientation with respect to its equilibrium position.
The physical mechanism behind this phenomenon is explained by considering the relative contributions of the active and polymeric stresses to the shear rate.
We emphasize that the nonlinearity in the model does not result in any known oscillator models when truncated at any specific order.

Oscillatory dynamics is ubiquitous in biological systems, from simple predator-prey models to the more complex descriptions of how neurons transmit electrical signal spikes, such as  the Hodgkin-Huxley model or its two-dimensional counterpart, the Fitzhugh-Nagumo model \cite{Hodgkin1952,Fitzhugh1955}. Indeed, recently, a sustained oscillatory behavior of bacteria in a viscoelastic fluid was reported and studied using both a full hydrodynamic active viscoelastic model~\cite{songliu2021visco}, and the Fitzhugh-Nagumo oscillator. Our two-dimensional model, motivated by concepts from sheared active nematics and in spite of only accounting for rotational relaxation and activity, captures the essential ingredients that will result in oscillatory behaviour: activity to generate a spontaneous flow and viscoelasticity that will generate sufficient amount of feedback. 
Our study thereby may provide a minimal framework to predict and analyze generic dynamical properties of biological systems, particularly those involving nematic active matter in a complex environment.

\section*{Acknowledgements}
A. D. acknowledges funding from the Novo Nordisk Foundation (Grant no. NNF18SA0035142), Villum Fonden (Grant no. 29476).

\section*{Competing Interests}
The authors declare no competing interests.

\section*{Authors contribution}
ELCMP, AD, and JMY conceptualized the study. ELCMP and HLT performed the analyses and simulations. All authors participated the preparation and submission of the manuscript.

\appendix
\section*{Verification of the trapping region}
Consider the domain $\Omega$ and the vector field that arises from Eqs.~\eqref{eq:dimensionless}. The following observations taken together mean that any trajectory beginning from the boundary remains confined within $\Omega$.

\begin{itemize}
    \item If $\theta_n=-\pi/4$, then, from Eq.~\ref{eq:dimensionless}a, $\cos2\theta_n=0$ so $\dot{\theta}_n=|\theta_n|/T_n>0$. Any trajectory starting from the left side of the box $\Omega$ will point to the right.
    \item If $\theta_n=\pi/4$, then, from Eq.~\ref{eq:dimensionless}a, $\dot{\theta}_n=-\theta_n/T_n<0$. Any trajectory starting from the right side of the box $\Omega$ will point to the left.
    \item If $\theta_p=-\pi/4$, then, from Eq.~\ref{eq:dimensionless}b, $\cos2\theta_p=0$ and so $\dot{\theta}_p=|\theta_p|/T_p>0$. Any trajectory starting from the bottom side of the box $\Omega$ will point upwards.
    \item If $\theta_p=\pi/4$, then, from Eq.~\ref{eq:dimensionless}b, $\dot{\theta}_p=-\theta_p/T_p<0$. Any trajectory starting from the top side of the box $\Omega$ will point downwards.
\end{itemize}   

We further remark that the same analysis but using \eqref{eq:fixedmod}a,b in the domain $\Omega$ results in the same observation that $\Omega$ is an external trapping region.

\bibliographystyle{unsrt}

\end{document}